\documentclass[prb,10pt,twocolumn,superscriptaddress,aps,floatfix]{revtex4-1}

\usepackage{chemformula} 
\usepackage[T1]{fontenc} 

\usepackage{graphicx}
\usepackage{url}
\usepackage[normalem]{ulem}
\usepackage{color}
\usepackage{mathtools,amssymb}
\usepackage{textcomp}
\usepackage{gensymb}
\usepackage{braket}

\begin{document}

\title{InAs/MoRe hybrid semiconductor/superconductor nanowire devices}

\author{Bilal Kousar}
\affiliation{Center for Quantum Devices, Niels Bohr Institute, University of Copenhagen, 2100 Copenhagen, Denmark}
\affiliation{Univ. Grenoble Alpes, CNRS, Institut Néel, 38000 Grenoble, France}

\author{Damon J. Carrad}
\email{damonc@dtu.dk}
\affiliation{Center for Quantum Devices, Niels Bohr Institute, University of Copenhagen, 2100 Copenhagen, Denmark}
\affiliation{Department of Energy Conversion and Storage, Technical University of Denmark, Fysikvej, Building 310, 2800 Kgs. Lyngby}

\author{Lukas Stampfer}
\affiliation{Center for Quantum Devices, Niels Bohr Institute, University of Copenhagen, 2100 Copenhagen, Denmark}

\author{Peter Krogstrup}
\affiliation{Center for Quantum Devices, Niels Bohr Institute, University of Copenhagen, 2100 Copenhagen,  Denmark}

\author{Jesper Nygård}
\affiliation{Center for Quantum Devices, Niels Bohr Institute, University of Copenhagen, 2100 Copenhagen,  Denmark}

\author{Thomas S. Jespersen}
\email{tsaje@dtu.dk}
\affiliation{Center for Quantum Devices, Niels Bohr Institute, University of Copenhagen, 2100 Copenhagen, Denmark}
\affiliation{Department of Energy Conversion and Storage, Technical University of Denmark, Fysikvej, Building 310, 2800 Kgs. Lyngby}



\begin{abstract} 
Implementing superconductors capable of proximity-inducing a large energy-gap in semiconductors in the presence of strong magnetic fields is a major goal towards applications of semiconductor/superconductor hybrid materials in future quantum information technologies. Here, we study the performance of devices consisting of InAs nanowires in electrical contact to molybdenum-rhenium (MoRe) superconducting alloys. The MoRe thin films exhibit transition temperatures $\sim 10 \, \mathrm K$ and critical fields exceeding $6 \, \mathrm T$. Normal/superconductor devices enabled tunnel spectroscopy of the corresponding induced superconductivity, which was maintained up to $\sim 10 \, \mathrm K$, and MoRe based Josephson devices exhibit supercurrents and multiple Andreev reflections. We determine an induced superconducting gap lower than expected from the transition temperature, and observe gap softening at finite magnetic field. These may be common features for hybrids based on large gap, type-II superconductors. The results encourage further development of MoRe-based hybrids.

\end{abstract}

\maketitle

The potential for building topologically protected quantum information processors from low-dimensional semiconductor-superconductor hybrid materials\cite{OregPRL10,LutchynPRL10} drives significant efforts across a range of fields including experimental and theoretical physics, device engineering, and material science.~\cite{lutchyn2018majorana} The main branch of experimental studies centers around hybrid semiconductor-superconductor devices employing a low band-gap, strong spin-orbit semiconductor such as InAs or InSb electrically connected to conventional superconductors such as aluminum\cite{AlbrechtNature16} or niobium.\cite{MourikScience12} Aluminum is a type-I superconductor with a transition temperature of $T_\mathrm{C} \sim 1.5 \, \mathrm K$ and low bulk critical magnetic field $B_\mathrm{C} \sim 10 \mathrm{mT}$. Although restricting to thin ($\leq 10$~nm) films and in-plane fields allows operating devices in high fields, obviating these restrictions motivates the development of hybrid materials with higher transition temperatures and bulk compatibility with magnetic fields.\cite{CarradAdvMat2020, pendharkar:2021, kanne:2021} Type-II superconducting films of niobium and niobium based alloys (NbT, NbTiN) are well developed for metallic superconducting devices and have been incorporated in semi/super hybrid materials and devices.\cite{MourikScience12, GulNL17, CarradAdvMat2020,  GharaviNanotech17,perla:2021} However, despite a bulk $T_\mathrm{C} \geq 9$~K, the induced superconducting gap in mesoscopic hybrids is often reduced to values comparable to those of Al\cite{grove-rasmussen:2009, MourikScience12, Gunel-JAP2012, CarradAdvMat2020}. This is typically attributed to the formation of Nb oxides, which are superconducting with a lower $T_\mathrm{C}^\mathrm{NbO} \approx 1.4\,\mathrm K$, metallic, or magnetic,\cite{hulm:1972} and prompts efforts to develop semiconductor/superconductor hybrids where the favourable bulk superconducting properties are retained at the mesoscale.\cite{CarradAdvMat2020, pendharkar:2021, kanne:2021} Since epitaxy between the highest single-element $T_\mathrm{C}$ superconductor and InAs has been demonstrated,\cite{kanne:2021} focus must now turn to compound superconductors to further increase $T_\mathrm{C}$.

Here we investigate type-II molybdenum-rhenium (MoRe) alloys with high $T_\mathrm{C} \approx 10 \, \mathrm K$ and $B_\mathrm C > 6 \, \mathrm T$ \cite{lerner:1966} as alternative materials for proximitizing InAs nanowires (NWs). The investigation of MoRe is further motivated by recent demonstrations of high quality MoRe superconducting resonators\cite{singh:2014} and bulk-like superconducting properties in very thin ($4 \, \mathrm{nm}$) films, even with a relatively large oxygen content $\sim 10\%$\cite{seleznev:2008a} suggesting tolerance to impurity incorporation exceeding that of Nb-based films. Finally, MoRe has been used as an electrode material in graphene,\cite{AmetScience16, calado_ballistic_2015, BorzenetsPRL16} carbon nanotubes, \cite{schneider:2012} and molecular \cite{gaudenzi:2015} superconducting devices, enabling operation in high magnetic field. However, a detailed spectroscopic study of the (sub-)gap properties of induced superconductivity from MoRe contacts has not been undertaken, and devices featuring semiconductors with large $g$-factors and strong spin-orbit coupling~\cite{OregPRL10, LutchynPRL10} are lacking. We therefore fabricated MoRe/InAs-nanowire/TiAu (S-NW-N) tunneling spectroscopy devices and MoRe/InAs-nanowire/MoRe (S-NW-S) Josephson devices and demonstrate crucial features of hybrid devices, including an induced superconducting gap present up to $T_\mathrm{C}$, sub-gap states and multiple Andreev reflections, with high yield. Unexpectedly, we find a reduced induced superconducting gap compared to that expected from the measured $T_\mathrm{C}$ of the contacts and a softening of the superconducting gap at low fields compared to $B_\mathrm{C}$ of the contacts. We discuss possible reasons for these observations and future directions.

\begin{figure}[t!]
\includegraphics[width=8.5 cm]{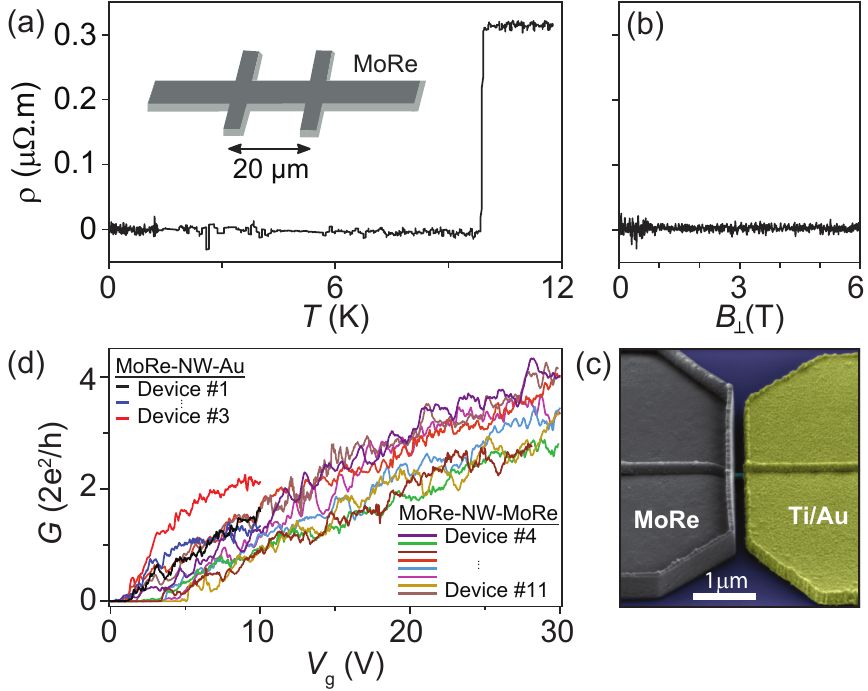}
\caption{(a,b) Dependence of the resistivity, $\rho$, on temperature and out-of-plane magnetic field, $B_\perp$, respectively, for a 250~nm thick MoRe film. The transition temperature $T_\mathrm{C}$ was 9.9~K, while the critical field was not reached up to 6~T. (c) SEM image of a typical S-NW-N device with MoRe (grey) and normal Ti/Au (yellow) electrodes. (d) Gate-dependence of the four terminal differential conductance $G = dI/dV$ measured at fixed source-drain bias $V_\mathrm{sd} = 4$~mV for eleven devices. In (b,d) the temperature was $15 \, \mathrm{mK}$}\label{Fig:1}
\end{figure}


MoRe thin films were deposited by DC magnetron sputtering from a commercially available target of Mo$_{0.5}$Re$_{0.5}$ using a power of 150~W under a 4~mTorr of Ar. The resulting deposition rate was of 11~nm/min and $250 \, \mathrm{nm}$ thick films were deposited on doped silicon substrates capped with $200 \, \mathrm{nm}$ of SiO$_2$ and patterning was achieved by conventional electron beam lithography and liftoff using a $\sim 350\, \mathrm{nm}$ thick poly methyl methacrylate resist. A 250~nm film thickness was chosen to ensure complete nanowire coverage, matching the typical thicknesses for Ti/Au contacts.\cite{KrogstrupNatMat15, ChangNatNano15, CarradAdvMat2020, kanne:2021} Devices for bulk characterization of the MoRe films were in the form of $3\ \mu \mathrm m$ wide MoRe strips measured in a four-terminal configuration with a $20\ \mu \mathrm m$ separation of the inner voltage probes (see sketch on Fig.\ 1a).

For hybrid devices, InAs NWs were grown by molecular beam epitaxy using Au catalyst particles following standard recipes\cite{KrogstrupNatMat15}. NWs have diameters $d_\mathrm{NW}  = 100-115 \, \mathrm{nm}$ and lengths $\sim 10~\mu \mathrm{m}$. Individual NWs were transferred from the growth substrate to the device substrate using a manual micro-manipulator and located with respect to a predefined alignment grid by optical microscopy. S-NW-N and S-NW-S devices have electrode separations of $95-130 \, \mathrm{nm}$ and $140-365 \, \mathrm{nm}$, respectively, and Ar$^+$ milling was performed before metal deposition with the aim of reducing native oxides and facilitating ohmic contacts. Figure \ref{Fig:1}c shows an SEM micrograph of a typical S-NW-N device. The electron density of the NW between the contacts was tuned by biasing the doped Si substrate at voltage $V_\mathrm g$, and each electrode was split into two to allow a pseudo four-terminal measurement of the differential conductance $G = \mathrm d I/ \mathrm d V_\mathrm{sd}$. As such, contributions from in-line resistances due to cryostat wiring/filtering are absent from measurements. Experiments were performed using standard lock-in techniques in a dilution refrigerator with a base temperature of $\sim 15 \, \mathrm{mK}$ and equipped with a 6-1-1 T vector magnet. In total, we measured four S-NW-N devices, fabricated on the same chip, three of which -- Devices 1-3 -- exhibited a superconducting gap in tunneling spectroscopy, and nine S-NW-S devices (Devices 4-12, fabricated on a second chip), all of which exhibited a supercurrent (Supporting Figs.\ S2 and S3). In the following we present results from one representative device of each class; data from remaining devices is presented in the Supporting Information. 

Figures 1a,b show the MoRe thin film resistivity, $\rho$, as a function of temperature and out-of-plane magnetic field $(B_\perp)$, respectively. The transition temperature was $T_\mathrm{C} = 9.9 \, \mathrm K$ and at base temperature, the film remained superconducting up to the highest possible fields in our setup, and thus the critical field $B^\mathrm{C}_\mathrm{MoRe}$ exceeds 6~T. These values are similar to recent studies of MoRe thin films having $T_\mathrm{C}$ of 8-10~K and $B_\mathrm{C}$ of 8~T \cite{AmetScience16, calado_ballistic_2015}, confirming the good quality and stoichiometry of our films.

Figure \ref{Fig:1}d shows $G(V_\mathrm g)$ for all devices. The measurements were performed with finite bias $V_\mathrm{sd} = 4 \, \mathrm{mV}$, outside the MoRe superconducting gap (see Supporting information for measurement schematics). The conductance increases with $V_\mathrm{g}$ as expected for $n$-type InAs and reaches $G \sim 6-8 e^2/h$ at the highest $V_\mathrm{g}$. The fluctuations in $G(V_\mathrm{g})$ are reproducible for each device and attributed to universal conductance fluctuations as expected under these conditions.\cite{jespersen:2006} The data at the highest $V_\mathrm{g}$ in Fig.~1d implies an upper bound on the contact resistance of a few k$\Omega$, typical for metal contacts to InAs nanowires. Due to screening by the grounded, semi-conformal contact metal, the gate is expected to predominately tune the nanowire segment between the contacts, however, a weak gate-dependence of the contact resistance cannot be ruled out, which may occur due to the generation of potential barriers that extend laterally beyond the contacts, as observed previously for InAs nanowires\cite{stormRealizingLateralWrapGated2012}. The device-to-device variation in conductance and threshold voltage is typical for these types of NWs, while the high yield (11/12) exceeds that typically obtained.\cite{CarradAdvMat2020,kanne:2021} The positive threshold voltage may be due to a Fermi level matching at the InAs/MoRe interface different from conventional InAs/Al or InAs/Au contacts, but other factors – charge traps, diameters, condition of cooldown, work-function etc. also influence the value. Positive threshold voltages as observed here are less common for InAs nanowires but key features of semiconductor/superconductor hybrids, e.g., hard induced gap, 2e-charging of islands, induced sub-gap states have also been demonstrated for wires with positive threshold voltages.\cite{kanne:2021}

\begin{figure}[t]
\includegraphics[width=8.5 cm]{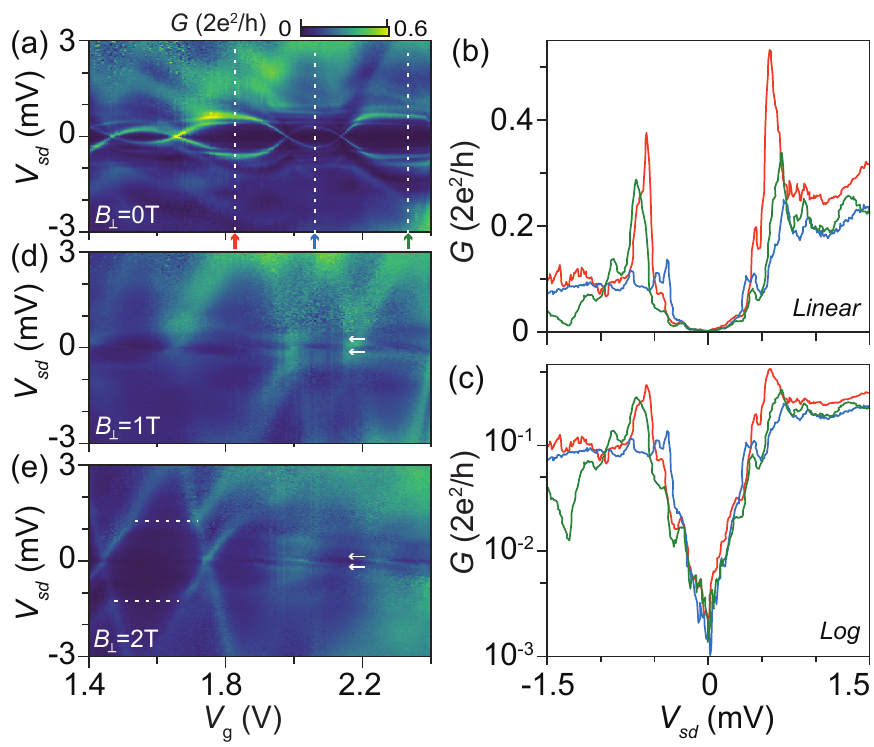}
\caption{(a) Measurement at $B=0$~T of $G$ vs. $V_\mathrm{sd}$ and $V_\mathrm g$ for S-NW-N Device 1 and $V_\mathrm g$ close to pinch off. (b,c) as (a) measured with finite perpendicular field $B_\perp = 1\, \mathrm T$ and $2\,\mathrm T$, respectively. Arrows indicate features at finite field associated with superconductivity of the MoRe leads. Dashed lines indicate co-tunneling features\cite{defranceschi-cotunneling}. (d,e) $G(V_\mathrm{sd})$ at fixed $V_\mathrm g$ as indicated in (a) plotted on linear and logarithmic scale, respectively.} \label{Fig:2}
\end{figure}


To investigate the induced superconducting properties we focus first on the S-NW-N devices. Figure \ref{Fig:2}a shows $G$ vs.\ $V_\mathrm{sd}$ and $V_\mathrm g$ close to pinch off. In this regime, quantum dots (QD) form in the junction,\cite{jespersen:2006} and when operated in Coulomb blockade, the QD acts as a tunnel barrier and the differential conductance, $G=dI/dV$, measures the density of states of the proximitized InAs\cite{ChangNatNano15}. The $V_\mathrm{g}$-dependent features observed at low bias are associated with Yu-Shiba-Rusinov (YSR) states arising from hybridisation between the QD levels and the superconducting contact\cite{deacon:2010}. Gate-independent features at higher bias can be associated with the superconducting gap, and/or bound states with an energy close to that of the gap. The induced superconducting gap is therefore estimated to be $\Delta^* \sim 0.9 \, \mathrm{meV}$, although better certainty may be obtained in future measurements by utilising a device geometry where the density of states underneath the superconductor can be depleted\cite{VaitiekenasPRL18gfactor}. Corresponding measurements for other devices yield $\Delta^* = 0.5 - 0.9$~meV (Supporting Fig.\ S1); we return to this variation below. The conventional Coulomb blockade (CB) diamonds are resolved in the corresponding measurements at finite perpendicular out-of-plane magnetic field, $(B_\perp)$, in Fig~2b,c, from which the charging energy $E_\mathrm{C} \sim 3 \, \mathrm{meV}$, level-spacing $\delta E \sim 1\, \mathrm{meV}$, and a level broadening of $\sim 0.1 \, \mathrm{meV}$ are estimated. The $g$-factor $g \sim 10.5$ was estimated from the Zeeman splitting of the excited state and corresponding co-tunneling lines around $V_\mathrm{g} = 1.6$~V (dotted lines in Fig. 2c).

An important metric for semiconductor/superconductor hybrids is the hardness of the induced gap\cite{TakeiPRL13, StanescuPRB14, ChangNatNano15} conventionally quantified by the ratio of $G_\mathrm S \equiv G(V_\mathrm{sd} = 0)$ to the out-of-gap conductance $G_\mathrm{N} = G(V_\mathrm{sd} > \Delta / e)$. Figures \ref{Fig:2}d,e show $G(V_\mathrm{sd})$ for $V_\mathrm g$ fixed at the centre of three consequtive CB diamonds as indicated in Fig.\ \ref{Fig:2}a where YSR-states are pushed towards the gap edge. From Figure \ref{Fig:2}e we find a $G_\mathrm{N} / G_\mathrm S \sim 100-150$, close to the values found for hybrids with epitaxial and impurity-free interfaces\cite{KrogstrupNatMat15,ChangNatNano15, CarradAdvMat2020, pendharkar:2021,kanne:2021}. However, contrary to the results of high-purity hybrids and metallic superconductors, the conductance in Fig.\ \ref{Fig:2}e is not uniformly reduced over the range of energies $eV_\mathrm{sd} < \Delta^*$ but rapidly increases with $V_\mathrm{sd}$. This 'V'-shape of $G(V_\mathrm{sd})$\cite{TakeiPRL13} is observed in all three charge states and can be observed at biases below $0.25 \, \mathrm{meV}$ far below any contributions to non-linearities of $G(V_\mathrm{sd})$ related to inelastic co-tunneling through the excited state at $\delta E$ with a broadening of $0.1 \, \mathrm{meV}$. Instead, we attribute the `V' shaped $G(V_\mathrm{sd})$ to a disorder-related continuum of interface states,~\cite{StanescuPRB14, TakeiPRL13, ChangNatNano15, KrogstrupNatMat15, HeedtNComm2021} possibly related to our fabrication scheme which includes mechanical Ar-ion milling of the InAs surface in contrast to the `U' shaped gap-profile of epitaxial electrodes\cite{ChangNatNano15,HeedtNComm2021}. In-situ deposition of MoRe,\cite{KrogstrupNatMat15,ChangNatNano15,CarradAdvMat2020,kanne:2021, HeedtNComm2021} or careful surface cleaning\cite{GazibegovicNature17, pendharkar:2021} would likely improve the sub-gap conductance profile. Note that the `U'-shape profile observed for hard-gap devices based on hybrids with similar $\Delta$, $E_\mathrm{C}$ and $\delta E \sim 1\, \mathrm{meV}$~\cite{kanne:2021,pendharkar:2021} supports the notion that the presence of interface inhomogeneities dominates the $G(V_\mathrm{sd})$ profile over possible secondary effects of, e.g., broadening due to Coulomb blockade at high biases, consistent with theory.\cite{TakeiPRL13}

We note that the magnitude of the induced gap $\Delta^* = 0.9 \, \mathrm{meV}$ is lower than the value $\Delta_\mathrm{bulk} = 1.75 k_\mathrm{B} T_\mathrm{C} = 1.4 \, \mathrm{meV}$ expected from the transition temperature $T_\mathrm{C} = 9.9 \, \mathrm K$ of the MoRe film (Fig.\ 1a). This is similar to MoRe/graphene devices showing a reduced $\Delta^* = 1.2$~meV.\cite{AmetScience16,BorzenetsPRL16} In analogy with results from Nb-based devices this could be a consequence of oxidized or contaminated layers at the non-ideal InAs/MoRe interface\cite{ChangNatNano15,GulNL17,TakeiPRL13,grove-rasmussen:2009}, defects in the superconductor\cite{cole:2016}, and/or carrier density-dependent interface transparency.\cite{Bubis_2017} Alternatively, the larger $\Delta_\mathrm{bulk}$ leads to a shorter coherence length -- $L^\mathrm{B}_\mathrm{C} = \hbar v_\mathrm{F}/\Delta_\mathrm{bulk}$ for ballistic transport and $L^\mathrm{D}_\mathrm{C} = \sqrt{\hbar v_\mathrm{F} l_\mathrm{e}/(2\Delta_{bulk})}$ for diffusive transport -- setting the scale beyond which $\Delta^*$ decays from the contact.\cite{courtois_1999,junger:2019,KjaergaardNatComm16} Taking a typical Fermi velocity $v_\mathrm{F} \sim 2 \times 10^5 \, \mathrm{m/s}$ and mean free paths $l_\mathrm{e}=50-200$~nm for InAs and $\Delta_\mathrm{bulk} = 1.4$~meV, we find $L^\mathrm{B}_\mathrm{C} = 100 \, \mathrm{nm}$ and $L^\mathrm{D}_\mathrm{C}=50-100$~nm, smaller than the electrode separation ($130\,\mathrm{nm}$ for Dev.\ 1). Thus, the gap measured by tunnel spectroscopy through the QD may be reduced compared value immediately below the MoRe contacts.\cite{KjaergaardNatComm16,junger:2019,courtois_1999,GueronPRL1996} We therefore posit that we primarily probe the induced proximity properties of the InAs adjacent to the MoRe contacts, which may account for the varying values of $\Delta^*$ obtained throughout. Advanced device architectures featuring in-situ deposition of a thin, few facet, impurity-free  MoRe layer would likely enable gate tuning of the proximitised region underneath the MoRe and exploration of its properties.~\cite{VaitiekenasPRL18gfactor} Relatedly, the bias-asymmetry in conductance in Fig. 2a likely arises from an asymmetric coupling to the leads.

\begin{figure*}[t]
\includegraphics[width=14 cm]{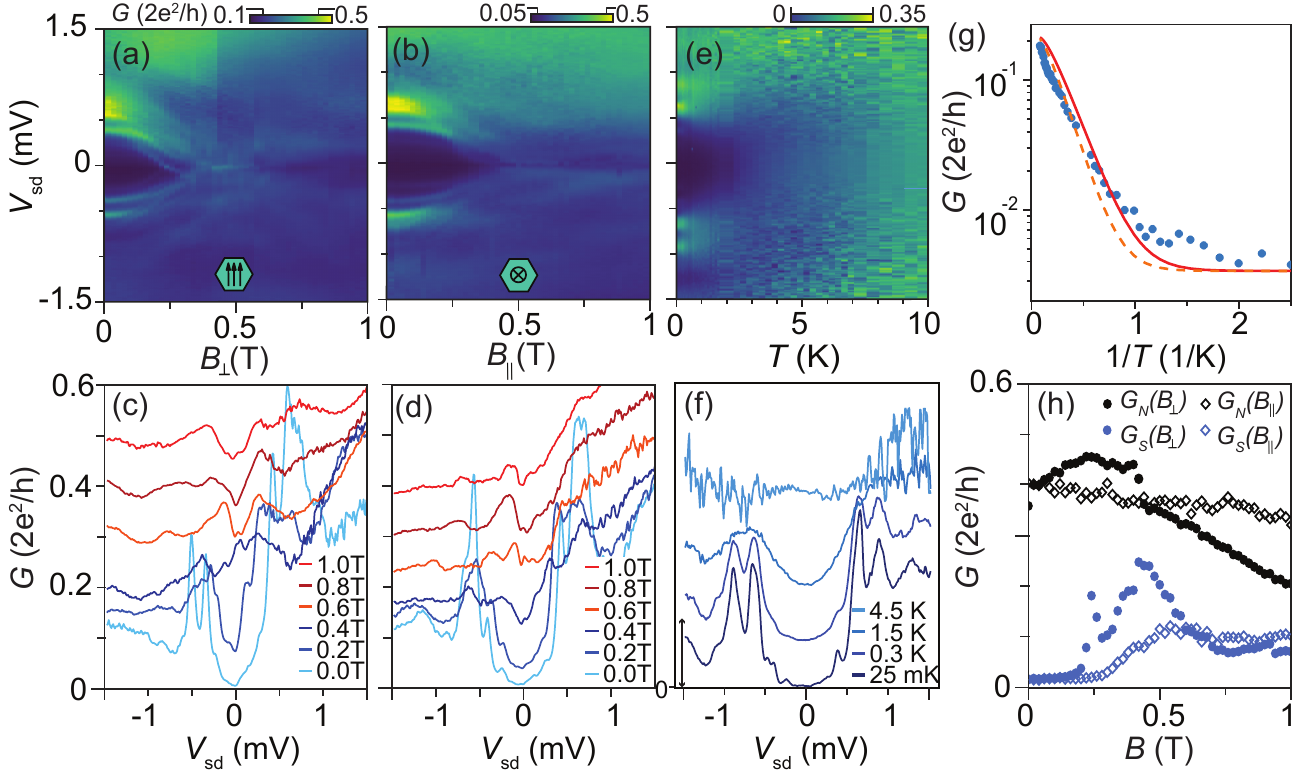}
\caption{(\label{Fig:3}a,b) Tunnelling spectroscopy as a function of perpendicular out-of-plane field and in-plane parallel field, respectively, with fixed $V_\mathrm{g} = 1.8 \, \mathrm{V}$ (red line in Fig.\ 2a). (c,d) Line traces of $G(V_\mathrm{sd})$ at fixed fields from Fig.\ 3a,b, respectively. (e,f) Same as panels (a) and (c) except measured as a function of temperature and for fixed $V_\mathrm g = 2.38 \, \mathrm V$ (green line in Fig.\ 2a). In (c,d,f) lines are off-set for clarity; in (f), vertical arrow corresponds to $0.2 \, 2\mathrm{e}^2/\mathrm h$. (g) Zero bias conductance $G_\mathrm{S}$ (points) as a function of inverse temperature. The solid red and dashed orange lines represent extrapolated fits performed up to $T=1.5$ and 3~K as described in the text. (h) $G_\mathrm{S}$ and $G_\mathrm{N}$ extracted from (a,b) as a function of $B_\perp$ (filled symbols), and $B_\parallel$ (open). The slight differences in each $G(V_\mathrm{sd})|_{B=0,T=15\mathrm{mK}}$ trace for the same $V_\mathrm{g}=1.8$~V arise from stochastic charge switches observed in e.g. (a) around $B=0.5$~T. These small differences do not affect the qualitative conclusions of this work, and all data in this figure were collected within the same Coulomb blockade diamond.} 
\end{figure*}

A key motivation for investigating MoRe as an electrode material is the compatibility of its superconducting phase with magnetic fields. Returning to Figs~\ref{Fig:2}b,c -- which show bias spectra with $B_\perp = 1\,\mathrm T$ and $B_\perp = 2\, \mathrm T$, respectively -- we observe that although the MoRe films are superconducting to $B^\mathrm{C}_\mathrm{bulk} > 6 \, \mathrm{T}$ (Fig.\ \ref{Fig:1}a), the superconductivity-related features around zero bias are strongly suppressed (white arrows) compared to the zero field case in Fig.\ \ref{Fig:1}a. The field dependence is detailed in Fig.\ \ref{Fig:3}a,b which show measurements of $G(V_\mathrm{sd})$ vs. $B_\perp$ and $B_\parallel$, respectively, at a fixed $V_\mathrm{g} = 1.8 \, \mathrm V$ within a CB diamond corresponding to even occupancy of the QD (red arrow in Fig.\ \ref{Fig:2}a). Here $B_\parallel$ is the in-plane field parallel to the axis of the NW. In both cases, states detach from the gap edge and fill the gap at $\sim 0.35 \, \mathrm T$ and $0.45 \, \mathrm T$ for $B_\perp$ and $B_\parallel$, respectively. This evolution is emphasized in Fig.\ \ref{Fig:3}c,d showing extracted $G(V_\mathrm{sd})$ for different fields, and in Fig.\ \ref{Fig:3}h, which shows $G_\mathrm S$ and $G_N$ as a function of $B$. A reduced gap remains visible and $G_\mathrm N / G_\mathrm S$ rapidly decreases for fields above $0.25 \, \mathrm T$. Similar observations have been reported for devices based on type-II superconductors Nb, NbTi, and NbTiN\cite{ZhangNatComm17,GulNatNano18} and associated to the formation of vortices in the electrodes, causing $\Delta$ to locally vanish thus softening the superconducting gap,\cite{LiuPRB17,GulNatNano18} and/or shifting spectral weight from the coherence peaks to energies within the gap.\cite{NikolaenkoArxhiv2020} We note, that for $B = 210 \, \mathrm{mT}$ the average vortex separation $\sqrt{4 \phi/3 B} $ reaches the NW diameter\cite{HessPRL1989} and thus $\Delta = 0$ vortex cores will be found proximal to the nanowire tunnel probe, softening the local density of states measurement.

\begin{figure}[!h]
\includegraphics[width=8 cm]{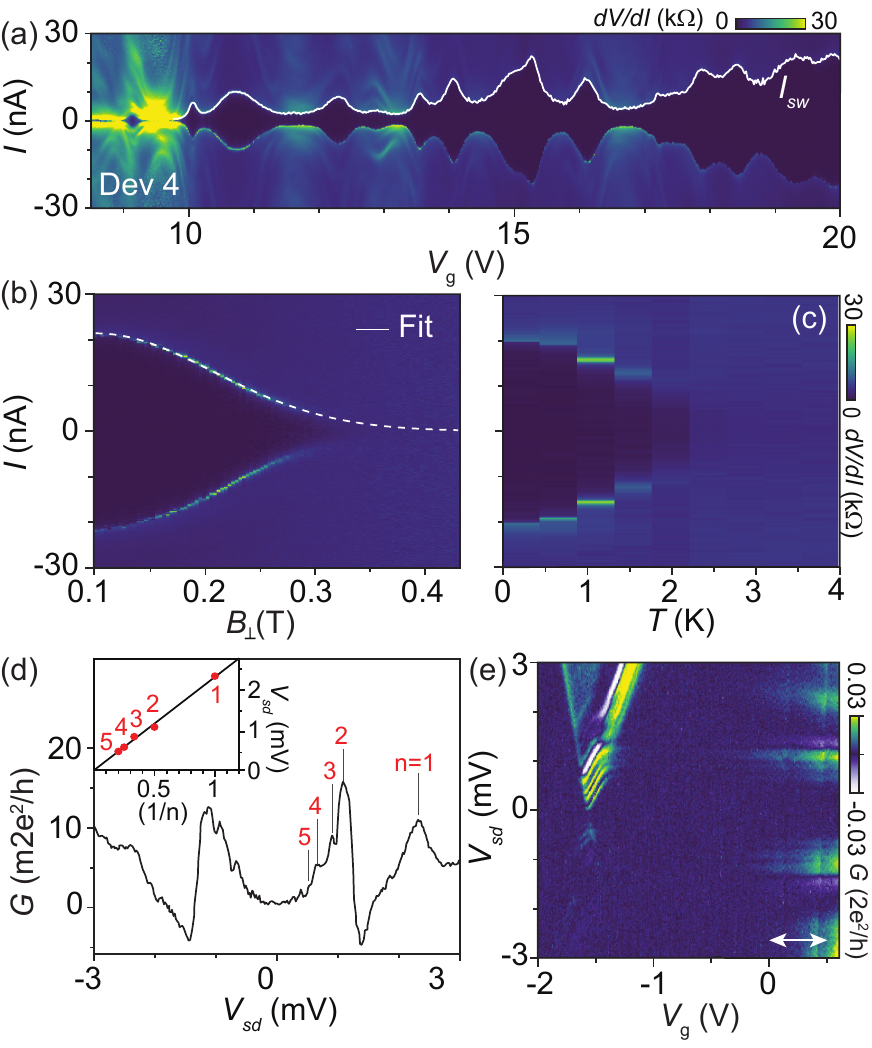}
\caption{\label{Fig:4} Properties of MoRe-NW-MoRe Josephson junctions. (a) Measurement of differential resistance $dV/dI$ vs.\ $I$ and $V_\mathrm{g}$, sweeping $I$ from negative to positive. Black region corresponds to the zero-resistance supercurrent state, and the transition to the dissipative normal state at the switching current $I_\mathrm {sw}$ is indicated. (b) Evolution of $I_\mathrm{sw}$ with $B_\perp$ for $V_\mathrm g = 20 \, \mathrm{V}$. The white lines shows a fit to the theory of a diffusive junction with high transmission contacts. (c) Corresponding temperature dependence with a supercurrent persisting up to $T \sim 2.5$~K~$<T_\mathrm{C}$. (d) Voltage bias measurement in the closed regime showing Coulomb blockade of a QD in the junction. Bias independent features in the CB diamond are associated with co-tunneling and multiple Andreev reflections. (e) $G$ vs $V_\mathrm{sd}$ trace averaged over $V_\mathrm{g} = 0 - 0.5$~V (white arrows in (d)). A harmonic series of MAR features $2\Delta/(en)$ is seen with assigned values of $n$ indicated. Inset: Extracted peak positions vs.\ $(1/n)$. The slope allows an estimate of the induced gap $\Delta^* = 1.1 \, \mathrm{meV}$.}
\end{figure}

Figure\ \ref{Fig:3}e,f,g show measurements similar to Fig.\ \ref{Fig:3}a,c,h except measured as a function of temperature and for $V_\mathrm g = 2.38 \, \mathrm V$ (green arrow in Fig.\ 2a). The coherence peaks weaken with temperature and the in-gap conductance increases markedly from $\sim 1 \, \mathrm K$ and reaches $G_\mathrm N$ at $\sim 10 \, \mathrm K$ (see Supporting Information Fig.~S5) consistent with $T_\mathrm C$ of the bulk MoRe film. Assuming $\Delta^*$ is temperature independent at $T \ll T^\mathrm{C}_\mathrm{bulk}$ the zero bias conductance was fitted to the conventional expression for the N-S tunnelling\cite{tinkham2004introduction} $\left.G_{\mathrm{S}}\right|_{V_{S D}=0}={G_{\mathrm{N}}}\sqrt{2 \pi \Delta^{*}/(k_{B} T)} \exp{(-\Delta^{*} / k_{B} T)}$. The solid red (dashed orange) line in Fig.\ \ref{Fig:3}g shows extrapolation of a fit performed up to $1.5 \, \mathrm K$ ($3 \, \mathrm K$), yielding $\Delta^* = 0.52 \, \mathrm{meV}$ ($0.62 \, \mathrm{meV}$). The dependence on fitting range, relatively poor fit at low $T$ (see Supporting Information for data plotted on a linear scale), and low values of $\Delta^*$ compared to those obtained from bias spectroscopy, suggests this fit may be affected by additional contributions to the conductance in addition to thermal excitation above $\Delta^*$, likely mediated by a high density of sub-gap states due to the non-ideal MoRe /InAs interface as discussed in connection with Fig. 2.\cite{ChangNatNano15} Future experiments involving high-purity interfaces\cite{ChangNatNano15, CarradAdvMat2020, GazibegovicNature17, kanne:2021, pendharkar:2021} may solve this issue.

Transport in S-NW-S Josephson devices where both electrodes are made from MoRe is governed by different phenomena than in S-NW-N devices: Dissipation-less supercurrents carried by Andreev bounds states dominate at zero bias, while multiple Andreev reflections dominate at finite bias. Figure 4a shows the differential resistance $dV/dI$ of S-NW-S device 4 as a function of $V_\mathrm g$ and the current $I$ through the device in a regime far from pinch-off ($G_\mathrm{N} \gtrsim 2~ e^{2}/h$). Here, $V$ is the measured voltage drop across the device, and measurements were performed in current-bias configuration with a $200$~M$\Omega$ bias resistor (see Supporting Information Figure S4). A zero-resistance state is observed at low currents with a switching current $I_\mathrm{sw}$ (white line) which decreases upon reducing $V_\mathrm{g}$ due to the depletion of the NW. 
In addition, $I_\mathrm{sw}$ exhibits modulations with $V_\mathrm g$ due to conductance fluctuations in the coherent InAs NW\cite{DohScience05}. In Fig.\ \ref{Fig:4}a, the current was swept from negative to positive and the minimal switching/retrapping hysteresis indicates overdamped junction dynamics and/or Joule heating.\cite{tinkham2004introduction} A maximal $I_\mathrm{C} = 23$~nA was found at $V_\mathrm{g} = 20$~V with a product $I_\mathrm{C} R_\mathrm{N} = 0.1 \mathrm{mV}  \ll \pi \Delta/(2e) = 2.2$~mV. The critical currents of nanowire JJs have been widely investigated and similar to the present findings, $I_\mathrm{C} R_\mathrm{N}$ is commonly found to considerably underestimate $\Delta$. A complete description of the underlying cause(s) is lacking, however disorder and/or inhomogeneity of materials and interfaces are likely an important factor\cite{DohScience05,khanHighlyTransparentGatable2020}.

Figure \ref{Fig:4}b shows the $dV/dI$ vs.\ $B_\perp$ at $V_\mathrm{g} = 20 \, \mathrm V$. The switching current $I_\mathrm{sw}$ decreases monotonically as expected from diffusive transport in narrow, long nanowire junctions where flux pick-up in the junction is suppressed.~\cite{hammer-PRB2007,cuevas-PRL2007} For a diffusive junction the decay is expected to follow the relation
$I_\mathrm{sw}(B_\perp) = I_\mathrm{sw}^0\ \mathrm{exp}\left[-\frac{0.145}{6} \left(\frac{\pi d_\mathrm{NW} L B_\perp}{\phi_0}\right)^2\right]$,\cite{hammer-PRB2007} with flux quantum $\phi_0 = h/2e$ and $I_\mathrm{sw}^0 = I_\mathrm{sw}(B_\perp = 0)$. The dashed line in Fig.\ \ref{Fig:4}c shows a fit to the measured $I_\mathrm{sw}$ with a fixed NW diameter $d_\mathrm{NW} = 115$~nm determined from SEM images of Dev.\ 4 and taking the length of the NW junction, $L$, as a free parameter. The fitted $L \sim 240$~nm is larger than the contact separation $\sim 175$~nm as previously observed for devices with a substantial semiconductor/superconductor overlap.\cite{PaajasteNanoLett.2015} This supports that the reduction of the measured induced gap, $\Delta^*$, in the S-NW-N may be influenced by a decay from the electrode to the QD position as discussed above. The temperature dependence of $I_\mathrm{sw}$ in Fig.\ 4c exhibits a decay typical for a diffusive junction and similar to the observed $I_\mathrm{sw}$ suppression below $B_\mathrm{C}^{bulk}$, complete suppression of $I_\mathrm{sw}$ occurs for $T\approx 3 \, \mathrm K <T_\mathrm{C}$ around the same temperature where $\Delta^*(T)$ vanishes -- cf.\ Fig.\ \ref{Fig:3}e,f.

Finally, we show in Fig.\ \ref{Fig:4}d,e bias spectroscopy in a regime close to pinch off  $V_\mathrm g \lesssim 0 \, \mathrm{V}$ where the device is dominated by a quantum dot in Coulomb blockade analogous to Fig.\ \ref{Fig:2}. For $V_\mathrm g$ and $V_\mathrm{sd}$ inside Coulomb blockade, sequential transport is suppressed, and the current is mediated by co-tunneling. As typical for this situation, the conductance exhibits gate independent structure at finite bias $eV_\mathrm{sd} = \pm 2\Delta^*/n, n= 1,2,3,\ldots$ due to the successive opening and closing of the allowed multiple Andreev reflection (MAR) processes. Since the MAR processes in Coulomb blockade regime occur across a randomly located quantum dot, the superconducting leads consist of the proximitized NW segments either side of the dot;\cite{FrederickPRB21} thus analysis of the MAR allows an alternative method for determining $\Delta^*$.\cite{DohScience05, RidderbosPRM19, AbayNL13} Figure 4e shows the extracted $G(V_\mathrm{sd})$ averaged over the $V_\mathrm g$-range indicated by a white arrow in Fig.\ \ref{Fig:4}d. A series of peaks are observed following the expected harmonic series with the indicated MAR order $n$. The inset shows the extracted peak positions as a function of $1/n$ and a linear fit gives $\Delta^* = 1.1 \, \mathrm{meV}$. This value is consistent with that obtained from N-NW-S tunnel spectroscopy given the observed sample to sample variation and close to the value obtained for MoRe/graphene devices.\cite{AmetScience16, BorzenetsPRL16} Continuous tunability between the `open' and Coulomb blockade regimes was seen in all devices (Supporting Figs.\ S2,S3).

To summarize, we have investigated the sputtered alloy molybdenum-rhenium Mo$_{0.5}$Re$_{0.5}$ as a superconducting contact material to InAs nanowires. Both MoRe-NW-N tunnel spectroscopy devices and MoRe-NW-MoRe Josephson devices were successfully fabricated and electrically characterized at cryogenic temperatures. The MoRe electrode films have a transition temperature of $9-10 \, \mathrm K$ corresponding to a superconducting gap $\Delta_\mathrm{bulk} = 1.4 \, \mathrm{meV}$, and the critical field exceeds $6 \, \mathrm{T}$. S-NW-N tunnel spectroscopy and analysis of the sub-gap structure in the conductance of S-NW-S devices shows an induced superconducting gap in the NW of $\sim 0.5-1.1 \, \mathrm{meV}$. This is lower than $\Delta_\mathrm{bulk}$ which we associate with disordered MoRe/NW interfaces or effects of a diffusive junction. Applying a magnetic field, we find that features associated with induced superconductivity are visible even for high perpendicular fields $B_\mathrm{\perp}>2\,\mathrm T$, however, the sub-gap conductance of the tunnel devices increases and the critical current of JJ devices decreases rapidly already within $500 \, \mathrm{mT}$. This finding is similar to previous studies of type-II superconductor/semiconductor hybrid devices. Although the nanowire exhibits a soft, `V'-shaped gap, we expect that the hardness could be improved by careful cleaning of the semiconductor interface,\cite{GazibegovicNature17} or by in-situ deposition after nanowire growth.\cite{KrogstrupNatMat15} MoRe thus appears a promising candidate to extend semiconductor/superconductor hybrids towards higher $T_\mathrm{C}$ operation, and potentially extended magnetic field ranges, if the role of vortices can be elucidated through, e.g., STM studies. In particular, we expect device performance to improve further if shadow epitaxy techniques\cite{CarradAdvMat2020,GazibegovicNature17,pendharkar:2021,khanHighlyTransparentGatable2020,HeedtNComm2021,KrizekNL17} can be applied to the sputter deposited films. In the short term, our results suggest MoRe as an alternative to Nb or NbTiN, offering simple processing, with superconducting properties less sensitive to oxide formation, and larger residual resistance ratios for RF applications.

\textbf{Acknowledgement}
The authors thank Shivendra Upadhyay for assistance with device fabrication. This research was supported by the Danish National Research Foundation, Microsoft Quantum Materials Lab, and by research grants from Villum Fonden (00013157), The Danish Council for Independent Research (7014-00132), and European Research Council (866158).

\section*{Competing interests} The authors declare no competing interests.

\section*{Supporting Information}
The Supporting Information contains circuit diagrams, additional tunnel spectroscopy data, additional analysis of the data in Fig. 3g and the raw data of Fig. 4d. This material is available free of charge via the internet at https://doi.org/10.1021/acs.nanolett.2c02532.

\providecommand{\latin}[1]{#1}
\makeatletter
\providecommand{\doi}
  {\begingroup\let\do\@makeother\dospecials
  \catcode`\{=1 \catcode`\}=2 \doi@aux}
\providecommand{\doi@aux}[1]{\endgroup\texttt{#1}}
\makeatother
\providecommand*\mcitethebibliography{\thebibliography}
\csname @ifundefined\endcsname{endmcitethebibliography}
  {\let\endmcitethebibliography\endthebibliography}{}


\begin{mcitethebibliography}{55}
\providecommand*\natexlab[1]{#1}
\providecommand*\mciteSetBstSublistMode[1]{}
\providecommand*\mciteSetBstMaxWidthForm[2]{}
\providecommand*\mciteBstWouldAddEndPuncttrue
  {\def\EndOfBibitem{\unskip.}}
\providecommand*\mciteBstWouldAddEndPunctfalse
  {\let\EndOfBibitem\relax}
\providecommand*\mciteSetBstMidEndSepPunct[3]{}
\providecommand*\mciteSetBstSublistLabelBeginEnd[3]{}
\providecommand*\EndOfBibitem{}
\mciteSetBstSublistMode{f}
\mciteSetBstMaxWidthForm{subitem}{(\alph{mcitesubitemcount})}
\mciteSetBstSublistLabelBeginEnd
  {\mcitemaxwidthsubitemform\space}
  {\relax}
  {\relax}

\bibitem[Oreg \latin{et~al.}(2010)Oreg, Refael, and {von Oppen}]{OregPRL10}
Oreg,~Y.; Refael,~G.; {von Oppen},~F. Helical {{Liquids}} and {{Majorana Bound
  States}} in {{Quantum Wires}}. \emph{Physical Review Letters} \textbf{2010},
  \emph{105}, 177002\relax
\mciteBstWouldAddEndPuncttrue
\mciteSetBstMidEndSepPunct{\mcitedefaultmidpunct}
{\mcitedefaultendpunct}{\mcitedefaultseppunct}\relax
\EndOfBibitem
\bibitem[Lutchyn \latin{et~al.}(2010)Lutchyn, Sau, and Das~Sarma]{LutchynPRL10}
Lutchyn,~R.~M.; Sau,~J.~D.; Das~Sarma,~S. Majorana {{Fermions}} and a
  {{Topological Phase Transition}} in {{Semiconductor}}-{{Superconductor
  Heterostructures}}. \emph{Physical Review Letters} \textbf{2010}, \emph{105},
  077001\relax
\mciteBstWouldAddEndPuncttrue
\mciteSetBstMidEndSepPunct{\mcitedefaultmidpunct}
{\mcitedefaultendpunct}{\mcitedefaultseppunct}\relax
\EndOfBibitem
\bibitem[Lutchyn \latin{et~al.}(2018)Lutchyn, Bakkers, Kouwenhoven, Krogstrup,
  Marcus, and Oreg]{lutchyn2018majorana}
Lutchyn,~R.~M.; Bakkers,~E.; Kouwenhoven,~L.~P.; Krogstrup,~P.; Marcus,~C.;
  Oreg,~Y. Majorana zero modes in superconductor--semiconductor
  heterostructures. \emph{Nature Reviews Materials} \textbf{2018}, \emph{3},
  52--68\relax
\mciteBstWouldAddEndPuncttrue
\mciteSetBstMidEndSepPunct{\mcitedefaultmidpunct}
{\mcitedefaultendpunct}{\mcitedefaultseppunct}\relax
\EndOfBibitem
\bibitem[Albrecht \latin{et~al.}(2016)Albrecht, Higginbotham, Madsen, Kuemmeth,
  Jespersen, Nyg{\aa}rd, Krogstrup, and Marcus]{AlbrechtNature16}
Albrecht,~S.~M.; Higginbotham,~A.~P.; Madsen,~M.; Kuemmeth,~F.;
  Jespersen,~T.~S.; Nyg{\aa}rd,~J.; Krogstrup,~P.; Marcus,~C.~M. Exponential
  Protection of Zero Modes in {{Majorana}} Islands. \emph{Nature}
  \textbf{2016}, \emph{531}, 206--209\relax
\mciteBstWouldAddEndPuncttrue
\mciteSetBstMidEndSepPunct{\mcitedefaultmidpunct}
{\mcitedefaultendpunct}{\mcitedefaultseppunct}\relax
\EndOfBibitem
\bibitem[Mourik \latin{et~al.}(2012)Mourik, Zuo, Frolov, Plissard, Bakkers, and
  Kouwenhoven]{MourikScience12}
Mourik,~V.; Zuo,~K.; Frolov,~S.~M.; Plissard,~S.~R.; Bakkers,~E. P. A.~M.;
  Kouwenhoven,~L.~P. Signatures of {{Majorana Fermions}} in {{Hybrid
  Superconductor}}-{{Semiconductor Nanowire Devices}}. \emph{Science}
  \textbf{2012}, \emph{336}, 1003--1007\relax
\mciteBstWouldAddEndPuncttrue
\mciteSetBstMidEndSepPunct{\mcitedefaultmidpunct}
{\mcitedefaultendpunct}{\mcitedefaultseppunct}\relax
\EndOfBibitem
\bibitem[Carrad \latin{et~al.}(2020)Carrad, Bjergfelt, Kanne, Aagesen, Krizek,
  Fiordaliso, Johnson, NygÃ¥rd, and Jespersen]{CarradAdvMat2020}
Carrad,~D.~J.; Bjergfelt,~M.; Kanne,~T.; Aagesen,~M.; Krizek,~F.;
  Fiordaliso,~E.~M.; Johnson,~E.; NygÃ¥rd,~J.; Jespersen,~T.~S. Shadow
  {Epitaxy} for {In} {Situ} {Growth} of {Generic}
  {Semiconductor}/{Superconductor} {Hybrids}. \emph{Advanced Materials}
  \textbf{2020}, \emph{32}, 1908411\relax
\mciteBstWouldAddEndPuncttrue
\mciteSetBstMidEndSepPunct{\mcitedefaultmidpunct}
{\mcitedefaultendpunct}{\mcitedefaultseppunct}\relax
\EndOfBibitem
\bibitem[Pendharkar \latin{et~al.}(2021)Pendharkar, Zhang, Wu, Zarassi, Zhang,
  Dempsey, Lee, Harrington, Badawy, Gazibegovic, {Op het Veld}, Rossi, Jung,
  Chen, Verheijen, Hocevar, Bakkers, Palmstr{\o}m, and Frolov]{pendharkar:2021}
Pendharkar,~M. \latin{et~al.}  Parity-Preserving and Magnetic Field\textendash
  Resilient Superconductivity in {{InSb}} Nanowires with {{Sn}} Shells.
  \emph{Science} \textbf{2021}, \emph{372}, 508--511\relax
\mciteBstWouldAddEndPuncttrue
\mciteSetBstMidEndSepPunct{\mcitedefaultmidpunct}
{\mcitedefaultendpunct}{\mcitedefaultseppunct}\relax
\EndOfBibitem
\bibitem[Kanne \latin{et~al.}(2021)Kanne, Marnauza, Olsteins, Carrad, Sestoft,
  {de Bruijckere}, Zeng, Johnson, Olsson, {Grove-Rasmussen}, and
  Nyg{\aa}rd]{kanne:2021}
Kanne,~T.; Marnauza,~M.; Olsteins,~D.; Carrad,~D.~J.; Sestoft,~J.~E.; {de
  Bruijckere},~J.; Zeng,~L.; Johnson,~E.; Olsson,~E.; {Grove-Rasmussen},~K.;
  Nyg{\aa}rd,~J. Epitaxial {{Pb}} on {{InAs}} Nanowires for Quantum Devices.
  \emph{Nat. Nanotechnol.} \textbf{2021}, \emph{16}, 776--781\relax
\mciteBstWouldAddEndPuncttrue
\mciteSetBstMidEndSepPunct{\mcitedefaultmidpunct}
{\mcitedefaultendpunct}{\mcitedefaultseppunct}\relax
\EndOfBibitem
\bibitem[G{\"u}l \latin{et~al.}(2017)G{\"u}l, Zhang, {de Vries}, {van Veen},
  Zuo, Mourik, {Conesa-Boj}, Nowak, {van Woerkom}, {Quintero-P{\'e}rez},
  Cassidy, Geresdi, Koelling, Car, Plissard, Bakkers, and Kouwenhoven]{GulNL17}
G{\"u}l,~{\"O}. \latin{et~al.}  Hard {{Superconducting Gap}} in {{InSb
  Nanowires}}. \emph{Nano Letters} \textbf{2017}, \emph{17}, 2690 -- 2696\relax
\mciteBstWouldAddEndPuncttrue
\mciteSetBstMidEndSepPunct{\mcitedefaultmidpunct}
{\mcitedefaultendpunct}{\mcitedefaultseppunct}\relax
\EndOfBibitem
\bibitem[Gharavi \latin{et~al.}(2017)Gharavi, Holloway, LaPierre, and
  Baugh]{GharaviNanotech17}
Gharavi,~K.; Holloway,~G.~W.; LaPierre,~R.~R.; Baugh,~J. Nb/{{InAs}} Nanowire
  Proximity Junctions from {{Josephson}} to Quantum Dot Regimes.
  \emph{Nanotechnology} \textbf{2017}, \emph{28}, 085202\relax
\mciteBstWouldAddEndPuncttrue
\mciteSetBstMidEndSepPunct{\mcitedefaultmidpunct}
{\mcitedefaultendpunct}{\mcitedefaultseppunct}\relax
\EndOfBibitem
\bibitem[Perla \latin{et~al.}(2021)Perla, Fonseka, Zellekens, Deacon, Han,
  K{\"o}lzer, M{\"o}rstedt, Bennemann, Espiari, Ishibashi, Gr{\"u}tzmacher,
  Sanchez, Lepsa, and Sch{\"a}pers]{perla:2021}
Perla,~P.; Fonseka,~H.~A.; Zellekens,~P.; Deacon,~R.; Han,~Y.; K{\"o}lzer,~J.;
  M{\"o}rstedt,~T.; Bennemann,~B.; Espiari,~A.; Ishibashi,~K.;
  Gr{\"u}tzmacher,~D.; Sanchez,~A.~M.; Lepsa,~M.~I.; Sch{\"a}pers,~T. Fully
  {\emph{in Situ}} {{Nb}}/{{InAs}}-Nanowire {{Josephson}} Junctions by
  Selective-Area Growth and Shadow Evaporation. \emph{Nanoscale Adv.}
  \textbf{2021}, \emph{3}, 1413--1421\relax
\mciteBstWouldAddEndPuncttrue
\mciteSetBstMidEndSepPunct{\mcitedefaultmidpunct}
{\mcitedefaultendpunct}{\mcitedefaultseppunct}\relax
\EndOfBibitem
\bibitem[Grove-Rasmussen \latin{et~al.}(2009)Grove-Rasmussen, JÃ¸rgensen,
  Andersen, Paaske, Jespersen, NygÃ¥rd, Flensberg, and
  Lindelof]{grove-rasmussen:2009}
Grove-Rasmussen,~K.; JÃ¸rgensen,~H.~I.; Andersen,~B.~M.; Paaske,~J.;
  Jespersen,~T.~S.; NygÃ¥rd,~J.; Flensberg,~K.; Lindelof,~P.~E.
  Superconductivity-Enhanced Bias Spectroscopy in Carbon Nanotube Quantum Dots.
  \emph{Phys. Rev. B} \textbf{2009}, \emph{79}, 134518\relax
\mciteBstWouldAddEndPuncttrue
\mciteSetBstMidEndSepPunct{\mcitedefaultmidpunct}
{\mcitedefaultendpunct}{\mcitedefaultseppunct}\relax
\EndOfBibitem
\bibitem[G{\"u}nel \latin{et~al.}(2012)G{\"u}nel, Batov, Hardtdegen, Sladek,
  Winden, Weis, Panaitov, Gr{\"u}tzmacher, and Sch{\"a}pers]{Gunel-JAP2012}
G{\"u}nel,~H.; Batov,~I.; Hardtdegen,~H.; Sladek,~K.; Winden,~A.; Weis,~K.;
  Panaitov,~G.; Gr{\"u}tzmacher,~D.; Sch{\"a}pers,~T. Supercurrent in
  {Nb/InAs-nanowire/Nb} {Josephson} junctions. \emph{Journal of Applied
  Physics} \textbf{2012}, \emph{112}, 034316\relax
\mciteBstWouldAddEndPuncttrue
\mciteSetBstMidEndSepPunct{\mcitedefaultmidpunct}
{\mcitedefaultendpunct}{\mcitedefaultseppunct}\relax
\EndOfBibitem
\bibitem[Hulm \latin{et~al.}(1972)Hulm, Jones, Hein, and Gibson]{hulm:1972}
Hulm,~J.~K.; Jones,~C.~K.; Hein,~R.~A.; Gibson,~J.~W. Superconductivity in the
  {{TiO}} and {{NbO}} Systems. \emph{J Low Temp Phys} \textbf{1972}, \emph{7},
  291--307\relax
\mciteBstWouldAddEndPuncttrue
\mciteSetBstMidEndSepPunct{\mcitedefaultmidpunct}
{\mcitedefaultendpunct}{\mcitedefaultseppunct}\relax
\EndOfBibitem
\bibitem[Lerner and Daunt(1966)Lerner, and Daunt]{lerner:1966}
Lerner,~E.; Daunt,~J.~G. Thermal and {{Electrical Conductivities}} of
  {{Mo}}-{{Re Alloys}} in the {{Superconducting}} and {{Normal States}}.
  \emph{Phys. Rev.} \textbf{1966}, \emph{142}, 251--258\relax
\mciteBstWouldAddEndPuncttrue
\mciteSetBstMidEndSepPunct{\mcitedefaultmidpunct}
{\mcitedefaultendpunct}{\mcitedefaultseppunct}\relax
\EndOfBibitem
\bibitem[Singh \latin{et~al.}(2014)Singh, Schneider, Bosman, Merkx, and
  Steele]{singh:2014}
Singh,~V.; Schneider,~B.~H.; Bosman,~S.~J.; Merkx,~E. P.~J.; Steele,~G.~A.
  Molybdenum-Rhenium Alloy Based High-{\emph{Q}} Superconducting Microwave
  Resonators. \emph{Appl. Phys. Lett.} \textbf{2014}, \emph{105}, 222601\relax
\mciteBstWouldAddEndPuncttrue
\mciteSetBstMidEndSepPunct{\mcitedefaultmidpunct}
{\mcitedefaultendpunct}{\mcitedefaultseppunct}\relax
\EndOfBibitem
\bibitem[Seleznev \latin{et~al.}(2008)Seleznev, Tarkhov, Voronov, Milostnaya,
  Lyakhno, Garbuz, Mikhailov, Zhigalina, and Gol'tsman]{seleznev:2008a}
Seleznev,~V.~A.; Tarkhov,~M.~A.; Voronov,~B.~M.; Milostnaya,~I.~I.;
  Lyakhno,~V.~Y.; Garbuz,~A.~S.; Mikhailov,~M.~Y.; Zhigalina,~O.~M.;
  Gol'tsman,~G.~N. Deposition and Characterization of Few-Nanometers-Thick
  Superconducting {{Mo}}-{{Re}} Films. \emph{Supercond. Sci. Technol.}
  \textbf{2008}, \emph{21}, 115006\relax
\mciteBstWouldAddEndPuncttrue
\mciteSetBstMidEndSepPunct{\mcitedefaultmidpunct}
{\mcitedefaultendpunct}{\mcitedefaultseppunct}\relax
\EndOfBibitem
\bibitem[Amet \latin{et~al.}(2016)Amet, Ke, Borzenets, Wang, Watanabe,
  Taniguchi, Deacon, Yamamoto, Bomze, Tarucha, and Finkelstein]{AmetScience16}
Amet,~F.; Ke,~C.~T.; Borzenets,~I.~V.; Wang,~J.; Watanabe,~K.; Taniguchi,~T.;
  Deacon,~R.~S.; Yamamoto,~M.; Bomze,~Y.; Tarucha,~S.; Finkelstein,~G.
  Supercurrent in the Quantum Hall Regime. \emph{Science} \textbf{2016},
  \emph{352}, 966--969\relax
\mciteBstWouldAddEndPuncttrue
\mciteSetBstMidEndSepPunct{\mcitedefaultmidpunct}
{\mcitedefaultendpunct}{\mcitedefaultseppunct}\relax
\EndOfBibitem
\bibitem[Calado \latin{et~al.}(2015)Calado, Goswami, Nanda, Diez, Akhmerov,
  Watanabe, Taniguchi, Klapwijk, and Vandersypen]{calado_ballistic_2015}
Calado,~V.~E.; Goswami,~S.; Nanda,~G.; Diez,~M.; Akhmerov,~A.~R.; Watanabe,~K.;
  Taniguchi,~T.; Klapwijk,~T.~M.; Vandersypen,~L. M.~K. Ballistic {Josephson}
  Junctions in Edge-Contacted Graphene. \emph{Nature Nanotechnology}
  \textbf{2015}, \emph{10}, 761--764\relax
\mciteBstWouldAddEndPuncttrue
\mciteSetBstMidEndSepPunct{\mcitedefaultmidpunct}
{\mcitedefaultendpunct}{\mcitedefaultseppunct}\relax
\EndOfBibitem
\bibitem[Borzenets \latin{et~al.}(2016)Borzenets, Amet, Ke, Draelos, Wei,
  Seredinski, Watanabe, Taniguchi, Bomze, Yamamoto, Tarucha, and
  Finkelstein]{BorzenetsPRL16}
Borzenets,~I.~V.; Amet,~F.; Ke,~C.~T.; Draelos,~A.~W.; Wei,~M.~T.;
  Seredinski,~A.; Watanabe,~K.; Taniguchi,~T.; Bomze,~Y.; Yamamoto,~M.;
  Tarucha,~S.; Finkelstein,~G. Ballistic Graphene Josephson Junctions from the
  Short to the Long Junction Regimes. \emph{Phys. Rev. Lett.} \textbf{2016},
  \emph{117}, 237002\relax
\mciteBstWouldAddEndPuncttrue
\mciteSetBstMidEndSepPunct{\mcitedefaultmidpunct}
{\mcitedefaultendpunct}{\mcitedefaultseppunct}\relax
\EndOfBibitem
\bibitem[Schneider \latin{et~al.}(2012)Schneider, Etaki, van~der Zant, and
  Steele]{schneider:2012}
Schneider,~B.~H.; Etaki,~S.; van~der Zant,~H. S.~J.; Steele,~G.~A. Coupling
  Carbon Nanotube Mechanics to a Superconducting Circuit. \emph{Scientific
  Reports} \textbf{2012}, \emph{2}, 599\relax
\mciteBstWouldAddEndPuncttrue
\mciteSetBstMidEndSepPunct{\mcitedefaultmidpunct}
{\mcitedefaultendpunct}{\mcitedefaultseppunct}\relax
\EndOfBibitem
\bibitem[Gaudenzi \latin{et~al.}(2015)Gaudenzi, Island, de~Bruijckere,
  BurzurÃƒÂ­, Klapwijk, and van~der Zant]{gaudenzi:2015}
Gaudenzi,~R.; Island,~J.~O.; de~Bruijckere,~J.; BurzurÃƒÂ­,~E.;
  Klapwijk,~T.~M.; van~der Zant,~H. S.~J. Superconducting Molybdenum-Rhenium
  Electrodes for Single-Molecule Transport Studies. \emph{Appl. Phys. Lett.}
  \textbf{2015}, \emph{106}, 222602\relax
\mciteBstWouldAddEndPuncttrue
\mciteSetBstMidEndSepPunct{\mcitedefaultmidpunct}
{\mcitedefaultendpunct}{\mcitedefaultseppunct}\relax
\EndOfBibitem
\bibitem[Krogstrup \latin{et~al.}(2015)Krogstrup, Ziino, Chang, Albrecht,
  Madsen, Johnson, Nyg{\aa}rd, Marcus, and Jespersen]{KrogstrupNatMat15}
Krogstrup,~P.; Ziino,~N. L.~B.; Chang,~W.; Albrecht,~S.~M.; Madsen,~M.~H.;
  Johnson,~E.; Nyg{\aa}rd,~J.; Marcus,~C.~M.; Jespersen,~T.~S. Epitaxy of
  Semiconductor\textendash{}Superconductor Nanowires. \emph{Nature Materials}
  \textbf{2015}, \emph{14}, 400--406\relax
\mciteBstWouldAddEndPuncttrue
\mciteSetBstMidEndSepPunct{\mcitedefaultmidpunct}
{\mcitedefaultendpunct}{\mcitedefaultseppunct}\relax
\EndOfBibitem
\bibitem[Chang \latin{et~al.}(2015)Chang, Albrecht, Jespersen, Kuemmeth,
  Krogstrup, Nyg{\aa}rd, and Marcus]{ChangNatNano15}
Chang,~W.; Albrecht,~S.~M.; Jespersen,~T.~S.; Kuemmeth,~F.; Krogstrup,~P.;
  Nyg{\aa}rd,~J.; Marcus,~C.~M. Hard Gap in Epitaxial
  Semiconductor\textendash{}Superconductor Nanowires. \emph{Nature
  Nanotechnology} \textbf{2015}, \emph{10}, 232--236\relax
\mciteBstWouldAddEndPuncttrue
\mciteSetBstMidEndSepPunct{\mcitedefaultmidpunct}
{\mcitedefaultendpunct}{\mcitedefaultseppunct}\relax
\EndOfBibitem
\bibitem[Jespersen \latin{et~al.}(2006)Jespersen, Aagesen, SÃ¸rensen, Lindelof,
  and NygÃ¥rd]{jespersen:2006}
Jespersen,~T.~S.; Aagesen,~M.; SÃ¸rensen,~C.; Lindelof,~P.~E.; NygÃ¥rd,~J.
  Kondo Physics in Tunable Semiconductor Nanowire Quantum Dots. \emph{Phys.
  Rev. B} \textbf{2006}, \emph{74}, 233304\relax
\mciteBstWouldAddEndPuncttrue
\mciteSetBstMidEndSepPunct{\mcitedefaultmidpunct}
{\mcitedefaultendpunct}{\mcitedefaultseppunct}\relax
\EndOfBibitem
\bibitem[Storm \latin{et~al.}(2012)Storm, Nylund, Samuelson, and
  Micolich]{stormRealizingLateralWrapGated2012}
Storm,~K.; Nylund,~G.; Samuelson,~L.; Micolich,~A.~P. Realizing {{Lateral
  Wrap-Gated Nanowire FETs}}: {{Controlling Gate Length}} with {{Chemistry
  Rather}} than {{Lithography}}. \emph{Nano Letters} \textbf{2012}, \emph{12},
  1--6\relax
\mciteBstWouldAddEndPuncttrue
\mciteSetBstMidEndSepPunct{\mcitedefaultmidpunct}
{\mcitedefaultendpunct}{\mcitedefaultseppunct}\relax
\EndOfBibitem
\bibitem[De~Franceschi \latin{et~al.}(2001)De~Franceschi, Sasaki, Elzerman,
  van~der Wiel, Tarucha, and Kouwenhoven]{defranceschi-cotunneling}
De~Franceschi,~S.; Sasaki,~S.; Elzerman,~J.~M.; van~der Wiel,~W.~G.;
  Tarucha,~S.; Kouwenhoven,~L.~P. Electron Cotunneling in a Semiconductor
  Quantum Dot. \emph{Phys. Rev. Lett.} \textbf{2001}, \emph{86}, 878--881\relax
\mciteBstWouldAddEndPuncttrue
\mciteSetBstMidEndSepPunct{\mcitedefaultmidpunct}
{\mcitedefaultendpunct}{\mcitedefaultseppunct}\relax
\EndOfBibitem
\bibitem[Deacon \latin{et~al.}(2010)Deacon, Tanaka, Oiwa, Sakano, Yoshida,
  Shibata, Hirakawa, and Tarucha]{deacon:2010}
Deacon,~R.~S.; Tanaka,~Y.; Oiwa,~A.; Sakano,~R.; Yoshida,~K.; Shibata,~K.;
  Hirakawa,~K.; Tarucha,~S. Tunneling {{Spectroscopy}} of {{Andreev Energy
  Levels}} in a {{Quantum Dot Coupled}} to a {{Superconductor}}. \emph{Phys.
  Rev. Lett.} \textbf{2010}, \emph{104}, 076805\relax
\mciteBstWouldAddEndPuncttrue
\mciteSetBstMidEndSepPunct{\mcitedefaultmidpunct}
{\mcitedefaultendpunct}{\mcitedefaultseppunct}\relax
\EndOfBibitem
\bibitem[Vaitiek{\.e}nas \latin{et~al.}(2018)Vaitiek{\.e}nas, Deng, Nyg{\aa}rd,
  Krogstrup, and Marcus]{VaitiekenasPRL18gfactor}
Vaitiek{\.e}nas,~S.; Deng,~M.-T.; Nyg{\aa}rd,~J.; Krogstrup,~P.; Marcus,~C.~M.
  Effective g-{{Factor}} of {{Subgap States}} in {{Hybrid Nanowires}}.
  \emph{Physical Review Letters} \textbf{2018}, \emph{121}, 037703\relax
\mciteBstWouldAddEndPuncttrue
\mciteSetBstMidEndSepPunct{\mcitedefaultmidpunct}
{\mcitedefaultendpunct}{\mcitedefaultseppunct}\relax
\EndOfBibitem
\bibitem[Takei \latin{et~al.}(2013)Takei, Fregoso, Hui, Lobos, and
  Sarma]{TakeiPRL13}
Takei,~S.; Fregoso,~B.~M.; Hui,~H.-Y.; Lobos,~A.~M.; Sarma,~S.~D. Soft
  {{Superconducting Gap}} in {{Semiconductor Majorana Nanowires}}.
  \emph{Physical Review Letters} \textbf{2013}, \emph{110}, 186803\relax
\mciteBstWouldAddEndPuncttrue
\mciteSetBstMidEndSepPunct{\mcitedefaultmidpunct}
{\mcitedefaultendpunct}{\mcitedefaultseppunct}\relax
\EndOfBibitem
\bibitem[Stanescu \latin{et~al.}(2014)Stanescu, Lutchyn, and
  Das~Sarma]{StanescuPRB14}
Stanescu,~T.~D.; Lutchyn,~R.~M.; Das~Sarma,~S. Soft Superconducting Gap in
  Semiconductor-Based {{Majorana}} Nanowires. \emph{Physical Review B}
  \textbf{2014}, \emph{90}, 085302\relax
\mciteBstWouldAddEndPuncttrue
\mciteSetBstMidEndSepPunct{\mcitedefaultmidpunct}
{\mcitedefaultendpunct}{\mcitedefaultseppunct}\relax
\EndOfBibitem
\bibitem[Heedt \latin{et~al.}(2021)Heedt, {Quintero-P{\'e}rez}, Borsoi,
  Fursina, {van Loo}, Mazur, Nowak, Ammerlaan, Li, Korneychuk, Shen, {van de
  Poll}, Badawy, Gazibegovic, {de Jong}, Aseev, {van Hoogdalem}, Bakkers, and
  Kouwenhoven]{HeedtNComm2021}
Heedt,~S. \latin{et~al.}  Shadow-Wall Lithography of Ballistic
  Superconductor\textendash Semiconductor Quantum Devices. \emph{Nature
  Communications} \textbf{2021}, \emph{12}, 4914\relax
\mciteBstWouldAddEndPuncttrue
\mciteSetBstMidEndSepPunct{\mcitedefaultmidpunct}
{\mcitedefaultendpunct}{\mcitedefaultseppunct}\relax
\EndOfBibitem
\bibitem[Gazibegovic \latin{et~al.}(2017)Gazibegovic, Car, Zhang, Balk, Logan,
  {de Moor}, Cassidy, Schmits, Xu, Wang, Krogstrup, {Op het Veld}, Zuo, Vos,
  Shen, Bouman, Shojaei, Pennachio, Lee, {van Veldhoven}, Koelling, Verheijen,
  Kouwenhoven, Palmstr{\o}m, and Bakkers]{GazibegovicNature17}
Gazibegovic,~S. \latin{et~al.}  Epitaxy of Advanced Nanowire Quantum Devices.
  \emph{Nature} \textbf{2017}, \emph{548}, 434--438\relax
\mciteBstWouldAddEndPuncttrue
\mciteSetBstMidEndSepPunct{\mcitedefaultmidpunct}
{\mcitedefaultendpunct}{\mcitedefaultseppunct}\relax
\EndOfBibitem
\bibitem[Cole \latin{et~al.}(2016)Cole, Sau, and Das~Sarma]{cole:2016}
Cole,~W.~S.; Sau,~J.~D.; Das~Sarma,~S. Proximity Effect and {{Majorana}} Bound
  States in Clean Semiconductor Nanowires Coupled to Disordered
  Superconductors. \emph{Phys. Rev. B} \textbf{2016}, \emph{94}, 140505\relax
\mciteBstWouldAddEndPuncttrue
\mciteSetBstMidEndSepPunct{\mcitedefaultmidpunct}
{\mcitedefaultendpunct}{\mcitedefaultseppunct}\relax
\EndOfBibitem
\bibitem[Bubis \latin{et~al.}(2017)Bubis, Denisov, Piatrusha, Batov, Khrapai,
  Becker, Treu, Ruhstorfer, and KoblmÃ¼ller]{Bubis_2017}
Bubis,~A.~V.; Denisov,~A.~O.; Piatrusha,~S.~U.; Batov,~I.~E.; Khrapai,~V.~S.;
  Becker,~J.; Treu,~J.; Ruhstorfer,~D.; KoblmÃ¼ller,~G. Proximity effect and
  interface transparency in Al/{InAs}-nanowire/Al diffusive junctions.
  \emph{Semiconductor Science and Technology} \textbf{2017}, \emph{32},
  094007\relax
\mciteBstWouldAddEndPuncttrue
\mciteSetBstMidEndSepPunct{\mcitedefaultmidpunct}
{\mcitedefaultendpunct}{\mcitedefaultseppunct}\relax
\EndOfBibitem
\bibitem[Courtois \latin{et~al.}(1999)Courtois, Gandit, Pannetier, and
  Mailly]{courtois_1999}
Courtois,~H.; Gandit,~P.; Pannetier,~B.; Mailly,~D. Long-range coherence and
  mesoscopic transport in Nâ€“S metallic structures. \emph{Superlattices and
  Microstructures} \textbf{1999}, \emph{25}, 721--732\relax
\mciteBstWouldAddEndPuncttrue
\mciteSetBstMidEndSepPunct{\mcitedefaultmidpunct}
{\mcitedefaultendpunct}{\mcitedefaultseppunct}\relax
\EndOfBibitem
\bibitem[J\"{u}nger \latin{et~al.}(2019)J\"{u}nger, Baumgartner, Delagrange,
  Chevallier, Lehmann, Nilsson, Dick, Thelander, and
  Sch\"{o}nenberger]{junger:2019}
J\"{u}nger,~C.; Baumgartner,~A.; Delagrange,~R.; Chevallier,~D.; Lehmann,~S.;
  Nilsson,~M.; Dick,~K.~A.; Thelander,~C.; Sch\"{o}nenberger,~C. Spectroscopy
  of the Superconducting Proximity Effect in Nanowires Using Integrated Quantum
  Dots. \emph{Communications Physics} \textbf{2019}, \emph{2}, 76\relax
\mciteBstWouldAddEndPuncttrue
\mciteSetBstMidEndSepPunct{\mcitedefaultmidpunct}
{\mcitedefaultendpunct}{\mcitedefaultseppunct}\relax
\EndOfBibitem
\bibitem[Kjaergaard \latin{et~al.}(2016)Kjaergaard, Nichele, Suominen, Nowak,
  Wimmer, Akhmerov, Folk, Flensberg, Shabani, Palmstr{\o}m, and
  Marcus]{KjaergaardNatComm16}
Kjaergaard,~M.; Nichele,~F.; Suominen,~H.~J.; Nowak,~M.~P.; Wimmer,~M.;
  Akhmerov,~A.~R.; Folk,~J.~A.; Flensberg,~K.; Shabani,~J.;
  Palmstr{\o}m,~C.~J.; Marcus,~C.~M. Quantized Conductance Doubling and Hard
  Gap in a Two-Dimensional Semiconductor\textendash{}Superconductor
  Heterostructure. \emph{Nature Communications} \textbf{2016}, \emph{7},
  12841\relax
\mciteBstWouldAddEndPuncttrue
\mciteSetBstMidEndSepPunct{\mcitedefaultmidpunct}
{\mcitedefaultendpunct}{\mcitedefaultseppunct}\relax
\EndOfBibitem
\bibitem[Gu\'eron \latin{et~al.}(1996)Gu\'eron, Pothier, Birge, Esteve, and
  Devoret]{GueronPRL1996}
Gu\'eron,~S.; Pothier,~H.; Birge,~N.~O.; Esteve,~D.; Devoret,~M.~H.
  Superconducting Proximity Effect Probed on a Mesoscopic Length Scale.
  \emph{Phys. Rev. Lett.} \textbf{1996}, \emph{77}, 3025--3028\relax
\mciteBstWouldAddEndPuncttrue
\mciteSetBstMidEndSepPunct{\mcitedefaultmidpunct}
{\mcitedefaultendpunct}{\mcitedefaultseppunct}\relax
\EndOfBibitem
\bibitem[Zhang \latin{et~al.}(2017)Zhang, G{\"u}l, {Conesa-Boj}, Nowak, Wimmer,
  Zuo, Mourik, de~Vries, van Veen, de~Moor, Bommer, van Woerkom, Car, Plissard,
  Bakkers, {Quintero-P{\'e}rez}, Cassidy, Koelling, Goswami, Watanabe,
  Taniguchi, and Kouwenhoven]{ZhangNatComm17}
Zhang,~H. \latin{et~al.}  Ballistic Superconductivity in Semiconductor
  Nanowires. \emph{Nature Communications} \textbf{2017}, \emph{8}, 16025\relax
\mciteBstWouldAddEndPuncttrue
\mciteSetBstMidEndSepPunct{\mcitedefaultmidpunct}
{\mcitedefaultendpunct}{\mcitedefaultseppunct}\relax
\EndOfBibitem
\bibitem[G{\"u}l \latin{et~al.}(2018)G{\"u}l, Zhang, Bommer, de~Moor, Car,
  Plissard, Bakkers, Geresdi, Watanabe, Taniguchi, and
  Kouwenhoven]{GulNatNano18}
G{\"u}l,~{\"O}.; Zhang,~H.; Bommer,~J. D.~S.; de~Moor,~M. W.~A.; Car,~D.;
  Plissard,~S.~R.; Bakkers,~E. P. A.~M.; Geresdi,~A.; Watanabe,~K.;
  Taniguchi,~T.; Kouwenhoven,~L.~P. Ballistic {{Majorana}} Nanowire Devices.
  \emph{Nature Nanotechnology} \textbf{2018}, \emph{13}, 192--197\relax
\mciteBstWouldAddEndPuncttrue
\mciteSetBstMidEndSepPunct{\mcitedefaultmidpunct}
{\mcitedefaultendpunct}{\mcitedefaultseppunct}\relax
\EndOfBibitem
\bibitem[Liu \latin{et~al.}(2017)Liu, Sau, and Das~Sarma]{LiuPRB17}
Liu,~C.-X.; Sau,~J.~D.; Das~Sarma,~S. Role of dissipation in realistic Majorana
  nanowires. \emph{Phys. Rev. B} \textbf{2017}, \emph{95}, 054502\relax
\mciteBstWouldAddEndPuncttrue
\mciteSetBstMidEndSepPunct{\mcitedefaultmidpunct}
{\mcitedefaultendpunct}{\mcitedefaultseppunct}\relax
\EndOfBibitem
\bibitem[Nikolaenko and Pientka(2021)Nikolaenko, and
  Pientka]{NikolaenkoArxhiv2020}
Nikolaenko,~A.; Pientka,~F. Topological superconductivity in proximity to
  type-II superconductors. \emph{Phys. Rev. B} \textbf{2021}, \emph{103},
  134503\relax
\mciteBstWouldAddEndPuncttrue
\mciteSetBstMidEndSepPunct{\mcitedefaultmidpunct}
{\mcitedefaultendpunct}{\mcitedefaultseppunct}\relax
\EndOfBibitem
\bibitem[Hess \latin{et~al.}(1989)Hess, Robinson, Dynes, Valles, and
  Waszczak]{HessPRL1989}
Hess,~H.~F.; Robinson,~R.~B.; Dynes,~R.~C.; Valles,~J.~M.; Waszczak,~J.~V.
  Scanning-{Tunneling}-{Microscope} {Observation} of the {Abrikosov} {Flux}
  {Lattice} and the {Density} of {States} near and inside a {Fluxoid}.
  \emph{Physical Review Letters} \textbf{1989}, \emph{62}, 214--216\relax
\mciteBstWouldAddEndPuncttrue
\mciteSetBstMidEndSepPunct{\mcitedefaultmidpunct}
{\mcitedefaultendpunct}{\mcitedefaultseppunct}\relax
\EndOfBibitem
\bibitem[Tinkham(2004)]{tinkham2004introduction}
Tinkham,~M. \emph{Introduction to Superconductivity}; Courier Corporation,
  2004\relax
\mciteBstWouldAddEndPuncttrue
\mciteSetBstMidEndSepPunct{\mcitedefaultmidpunct}
{\mcitedefaultendpunct}{\mcitedefaultseppunct}\relax
\EndOfBibitem
\bibitem[Doh \latin{et~al.}(2005)Doh, van Dam, Roest, Bakkers, Kouwenhoven, and
  Franceschi]{DohScience05}
Doh,~Y.-J.; van Dam,~J.~A.; Roest,~A.~L.; Bakkers,~E. P. A.~M.;
  Kouwenhoven,~L.~P.; Franceschi,~S.~D. Tunable {{Supercurrent Through
  Semiconductor Nanowires}}. \emph{Science} \textbf{2005}, \emph{309},
  272--275\relax
\mciteBstWouldAddEndPuncttrue
\mciteSetBstMidEndSepPunct{\mcitedefaultmidpunct}
{\mcitedefaultendpunct}{\mcitedefaultseppunct}\relax
\EndOfBibitem
\bibitem[Khan \latin{et~al.}(2020)Khan, Lampadaris, Cui, Stampfer, Liu, Pauka,
  Cachaza, Fiordaliso, Kang, Korneychuk, Mutas, Sestoft, Krizek, Tanta,
  Cassidy, Jespersen, and Krogstrup]{khanHighlyTransparentGatable2020}
Khan,~S.~A. \latin{et~al.}  Highly {{Transparent Gatable Superconducting Shadow
  Junctions}}. \emph{ACS Nano} \textbf{2020}, \emph{14}, 14605\relax
\mciteBstWouldAddEndPuncttrue
\mciteSetBstMidEndSepPunct{\mcitedefaultmidpunct}
{\mcitedefaultendpunct}{\mcitedefaultseppunct}\relax
\EndOfBibitem
\bibitem[Hammer \latin{et~al.}(2007)Hammer, Cuevas, Bergeret, and
  Belzig]{hammer-PRB2007}
Hammer,~J.; Cuevas,~J.~C.; Bergeret,~F.; Belzig,~W. Density of States and
  Supercurrent in Diffusive {SNS} Junctions: Roles of Nonideal Interfaces and
  Spin-Flip Scattering. \emph{Physical Review B} \textbf{2007}, \emph{76},
  064514\relax
\mciteBstWouldAddEndPuncttrue
\mciteSetBstMidEndSepPunct{\mcitedefaultmidpunct}
{\mcitedefaultendpunct}{\mcitedefaultseppunct}\relax
\EndOfBibitem
\bibitem[Cuevas and Bergeret(2007)Cuevas, and Bergeret]{cuevas-PRL2007}
Cuevas,~J.; Bergeret,~F. Magnetic interference patterns and vortices in
  diffusive {SNS} junctions. \emph{Physical Review Letters} \textbf{2007},
  \emph{99}, 217002\relax
\mciteBstWouldAddEndPuncttrue
\mciteSetBstMidEndSepPunct{\mcitedefaultmidpunct}
{\mcitedefaultendpunct}{\mcitedefaultseppunct}\relax
\EndOfBibitem
\bibitem[Paajaste \latin{et~al.}(2015)Paajaste, Amado, Roddaro, Bergeret,
  Ercolani, Sorba, and Giazotto]{PaajasteNanoLett.2015}
Paajaste,~J.; Amado,~M.; Roddaro,~S.; Bergeret,~F.~S.; Ercolani,~D.; Sorba,~L.;
  Giazotto,~F. Pb/{{InAs Nanowire Josephson Junction}} with {{High Critical
  Current}} and {{Magnetic Flux Focusing}}. \emph{Nano Letters} \textbf{2015},
  \emph{15}, 1803--1808\relax
\mciteBstWouldAddEndPuncttrue
\mciteSetBstMidEndSepPunct{\mcitedefaultmidpunct}
{\mcitedefaultendpunct}{\mcitedefaultseppunct}\relax
\EndOfBibitem
\bibitem[Thomas \latin{et~al.}(2021)Thomas, Nilsson, Ciaccia, J\"unger, Rossi,
  Zannier, Sorba, Baumgartner, and Sch\"onenberger]{FrederickPRB21}
Thomas,~F.~S.; Nilsson,~M.; Ciaccia,~C.; J\"unger,~C.; Rossi,~F.; Zannier,~V.;
  Sorba,~L.; Baumgartner,~A.; Sch\"onenberger,~C. Spectroscopy of the local
  density of states in nanowires using integrated quantum dots. \emph{Phys.
  Rev. B} \textbf{2021}, \emph{104}, 115415\relax
\mciteBstWouldAddEndPuncttrue
\mciteSetBstMidEndSepPunct{\mcitedefaultmidpunct}
{\mcitedefaultendpunct}{\mcitedefaultseppunct}\relax
\EndOfBibitem
\bibitem[Ridderbos \latin{et~al.}(2019)Ridderbos, Brauns, Li, Bakkers,
  Brinkman, van~der Wiel, and Zwanenburg]{RidderbosPRM19}
Ridderbos,~J.; Brauns,~M.; Li,~A.; Bakkers,~E. P. A.~M.; Brinkman,~A.; van~der
  Wiel,~W.~G.; Zwanenburg,~F.~A. Multiple Andreev Reflections and Shapiro Steps
  in a Ge-Si Nanowire Josephson Junction. \emph{Phys. Rev. Materials}
  \textbf{2019}, \emph{3}, 084803\relax
\mciteBstWouldAddEndPuncttrue
\mciteSetBstMidEndSepPunct{\mcitedefaultmidpunct}
{\mcitedefaultendpunct}{\mcitedefaultseppunct}\relax
\EndOfBibitem
\bibitem[Abay \latin{et~al.}(2013)Abay, Persson, Nilsson, Xu, Fogelstrom,
  Shumeiko, and Delsing]{AbayNL13}
Abay,~S.; Persson,~D.; Nilsson,~H.; Xu,~H.~Q.; Fogelström,~M.; Shumeiko,~V.;
  Delsing,~P. Quantized {Conductance} and {Its} {Correlation} to the
  {Supercurrent} in a {Nanowire} {Connected} to {Superconductors}. \emph{Nano
  Letters} \textbf{2013}, \emph{13}, 3614--3617\relax
\mciteBstWouldAddEndPuncttrue
\mciteSetBstMidEndSepPunct{\mcitedefaultmidpunct}
{\mcitedefaultendpunct}{\mcitedefaultseppunct}\relax
\EndOfBibitem
\bibitem[Krizek \latin{et~al.}(2017)Krizek, Kanne, Razmadze, Johnson,
  Nyg{\aa}rd, Marcus, and Krogstrup]{KrizekNL17}
Krizek,~F.; Kanne,~T.; Razmadze,~D.; Johnson,~E.; Nyg{\aa}rd,~J.;
  Marcus,~C.~M.; Krogstrup,~P. Growth of {{InAs Wurtzite Nanocrosses}} from
  {{Hexagonal}} and {{Cubic Basis}}. \emph{Nano Letters} \textbf{2017},
  \emph{17}, 6090--6096\relax
\mciteBstWouldAddEndPuncttrue
\mciteSetBstMidEndSepPunct{\mcitedefaultmidpunct}
{\mcitedefaultendpunct}{\mcitedefaultseppunct}\relax
\EndOfBibitem
\end{mcitethebibliography}
\end{document}